# 4: TELEROBOTIC MARS MISSION FOR LAVA TUBE EXPLORATION AND EXAMINATION OF LIFE

**Hanjo Schnellbaecher**
Hanjo.Schnellbaecher@TUDSaT.space
**Florian Dufresne**
Dufresne.Florian@gmail.com
**Tommy Nilsson**
Tommy.Nilsson@ESA.int
**Leonie Becker**
Leonie.Bensch@DLR.de
**Oliver Bensch**
Oliver.Bensch@DLR.de
**Enrico Guerra**
EnricoGuerra@outlook.com
**Wafa Sadri**
Wafa.Sadri@ESA.int
**Vanessa Neumann**
Vanessa.Neumann@TUDSaT.space

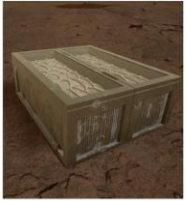
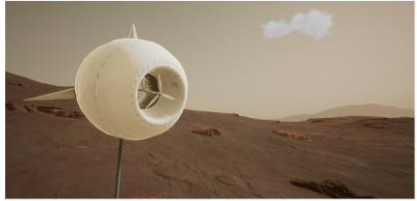

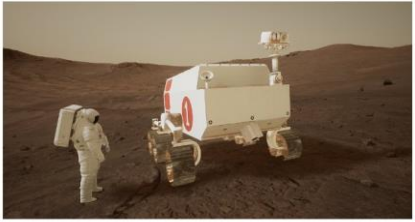
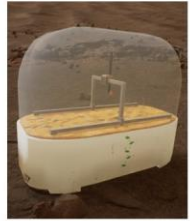

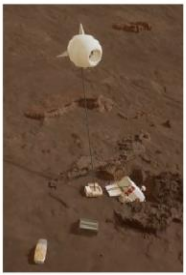
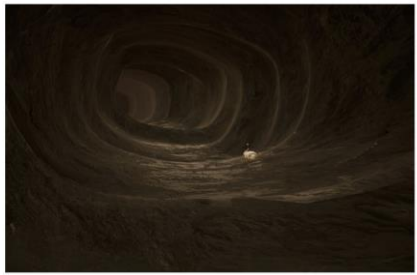

## AIM AND GENERAL PHILOSOPHY

The general profile and overarching goal of our proposed mission is to pioneer potentially highly beneficial, or even vital, and cost-effective techniques for the future human colonization of Mars. Adopting radically new and disruptive solutions untested in the Martian context, our approach is one of high risk and high reward. The real possibility of such a solution failing has prompted us to base our mission architecture around a rover carrying a set of 6 distinct experimental payloads, each capable of operating independently on the others, thus substantially increasing the chances of the mission yielding some valuable findings. At the same time, we sought to exploit available synergies by assembling a combination of payloads that would together form a coherent experimental ecosystem, with each payload providing potential value to the others. Apart from providing such a testbed for evaluation of novel technological solutions, another aim of our proposed mission is to help generate scientific know-how enhancing our understanding of the Red Planet.

Mars has been attracting scientific attention predominantly as the most likely planet to provide direct indication of life beyond Earth [1] as well as for its potential habitability [2]. While several robotic missions seeking to find signs of Martian life have already taken place (e.g., Curiosity), substantial areas of the Martian landscape remain unexplored. Chiefly, research indicates that lava tubes on Mars might provide conditions particularly conducive to life, due to stable temperatures and shielding from radiation [3]. Of equal interest is the exploration of conditions that might support life on Mars in the future. Developing reliable strategies for plant growth, for instance, will likely prove crucial for future Martian outposts. By way of example, studies on Earth have shown that certain species of fungi can thrive in extreme environments and even develop resilience to high levels of radiation [4]. Our ability to understand and take advantage of such opportunities might prove indispensable for humanity's future colonization of Mars.

To this end, our mission takes aim at the Nili-Fossae region, rich in natural resources (and carbonates in particular), past water repositories and signs of volcanic activity. With our proposed experimental payloads, we intend to explore existing lava -tubes, search for signs of past life and assess their potentially valuable geological features for future base building. We will evaluate biomatter in the form of plants and fungi as possible food and base-building materials respectively. Finally, we seek to explore a variety of novel power generation techniques using the Martian atmosphere and gravity. As detailed throughout the remainder of this report, this assemblage of experimental payloads, then, constitutes the backbone of our proposed telerobotic mission to Mars.

## LOCATION

Our search for a preferred landing site was primarily guided by two goals: maximizing the potential for scientific yield and minimizing environmental risks posed to the mission.

In terms of scientific yield, the regions on Mars deemed to be of greatest interest are the ones believed to once have contained water, hence being more likely to have preserved signs of past life [5]. A range of such sites scattered across the planet has been identified through previous research [6].

In terms of risk mitigation, the Martian southern hemisphere is mountainous, covered by complex terrain, making it difficult for rovers (or any other surface technology) to safely move around. In contrast, the northern hemisphere is considerably flatter, consisting mostly of lowlands. This makes navigation both easier and safer. Prior rover missions were targeting the northern hemisphere for this reason.

Drawing on such past missions and existing research, we determined that the Nili Fossae region would be the most suitable landing site. Located in the northern hemisphere, not only is this region comparatively safe and accessible for exploratory activities, but it is likewise believed to have once been rich in water [7] and consequently represents a suitable location for the search of life. Furthermore, as elaborated in the following subsection, evidence of past volcanic activity suggests that Nili Fossae might likewise be home to unexplored lava tubes [8].

### Lava Tube Exploration

Future human exploration missions to Mars will inevitably entail significant challenges, ranging from high radiation levels, micrometeoroids to general harsh climate conditions on the Martian surface [9]. One possible solution to mitigate these adverse environmental conditions on the red planet is the use of lava tubes that have formed naturally beneath the surface as shelter for human exploration. In this regard, [9] concluded that "Caves may be among the only structures on Mars that offer long-term protection from such hazards." (p. 1).

The structures do not only protect potential astronauts from radiations, but the ambient temperature inside lava tube structures remains constant, which would yield important advantages in comparison to highly fluctuating and extreme temperatures on the Martian surface [3].

| Scenario | Exposure Time | Cumulative Dose (mSV) |
|---|---|---|
| Surface Mission | 24hr/day | 14.795 |
| Cave Habitat | 24hr/day | 0.012 |

Table 1: Radiation Dose Comparison Mars Surface vs. Cave Habitat. Reprinted from [8].

Lava tubes were formed by the flow of low viscosity basaltic lava during the eruption of a non-explosive volcano [10]. Due to the lower gravity conditions, the diameter of lava tubes on Mars is thought to be several hundreds of meters wide [8], with walls tens of meters thick [11].

Yet, the exploration of lava tubes would not only be important with regards to future human settlements, but also from a research perspective, as life forms would also be protected inside the caves from hazardous conditions on the Martian surface [3]. Importantly, water resources could have been trapped and protected by the environment inside the tubes [11]. As lava tubes are generally located under the Martian surface, they would also yield a unique opportunity to scientifically explore minerals and material samples that are not accessible from the surface. The analysis of material probes that have been protected inside lava tubes could therefore also yield insights into the developmental history of Mars in terms of volcanic and thermal activity [3]. Similar missions have also already been investigated: For instance, the NASA-funded "BRAILLE" (Biologic and Resource Analog Investigations in Low Light Environments) project explores potential use of analog cave environments as a training ground for future lava tube missions. Simulation of scenarios involving search for lifeforms as well as life-sustaining minerals is of particular interest.

Based on these findings, we propose that one of our mission goals is the exploration of lava tubes to locate possible water resources, minerals and possibly even lifeforms that have been protected inside the structures, as this is a highly relevant research area with a considerable potential for scientific yield and value.

### Lava Tube Locations

Several different locations for lava tube entrances have been identified and mapped. In this regard, for instance, seven skylight entrances could be observed around the Arsia Mons using the orbiting Thermal Emission Imaging System (THEMIS) of the Mars Odyssey orbiter, which uses infrared imaging to detect subsurface structures, or the HiRISE camera on the Mars Reconnaissance Orbiter, which operates in visible wavelengths [12], [9]. As for the Nili Fossae region - the proposed landing site for our mission - it is located in the Volcanic region of Syrtis major and consequently considered likely to be the home of lava tubes under its surface [8]. Actual skylight entrances of lava tubes around Nili Fossae have yet to be spotted using data from the Mars Odyssey Orbiter though. As our mission is oriented towards scouting and exploration, the risk of placing our mission in Nili Fossae without finding any lava-tubes may be acceptable, especially considering that most of the technologies and experiments would still provide critical data (e.g. Greenhouse (see sec. 8), wind balloon (see sec. 5) ...). A location like Arisa Mons, for which lava tube entrances have already been spotted, may also serve as a backup location in case the Mars Reconnaissance orbiter would fail at finding lava tubes entrances near Nili Fossae within the 10-year window that remains till mission launch.

The planned payloads for the telerobotic rover mission are proposed and explained in detail in the following sections. In this regard, each subsystem's description will include an overview of the engineering design, as well as the value for exploration preparation and scientific return. In addition, estimated costs for each payload and a schedule will be provided.

Cost estimations for development and integration are based on the average NASA salary of 124.363$ [13]. As defined by NASA's Procedural Requirements NPR 7120.5F [14], the payloads were divided into subsystems according to NASA's Work Breakdown Structure (level 3) [15] and costs were calculated for each subsystem individually. Total costs for all payloads (WBS level 2) for the Project Life-Cycle Cost Estimate (LCCE) are shown in section 11.

The rover platform design will be presented after the presentation of six different payload subsystems, followed by a description of an overall mission schedule and timeline based on a project life-cycle model.

## PAYLOAD 1: GROUND PENETRATING RADAR & CAMERA SYSTEM

### Engineering Design
The proposed rover will use the Mastcam-Z, that is also attached to the Perseverance rover that was launched during the Mars 2020 mission. As was the case in the Mars 2020 mission, the camera will be attached to the body of the main rover [16]. For a detailed description of the system characteristics associated with the Mastcam-Z system, please refer to [16]. We plan to reuse the Mastcam-Z camera, despite its 2-megapixel camera, since it was constructed with the intent to deliver as much information as possible, with limited bandwidth available.

The ground penetrating radar system will be based on the RIMFAX system that was proposed by NASA [18] and utilized for the Perseverance rover mission. The technology can scan up to a depth of around 10m. Please refer to table 2 for a detailed description of all the system characteristics.

| Location | The radar antenna is on the lower rear of the rover |
|---|---|
| Mass | Less than 6.6 pounds (3 kilograms) |
| Power | 5 to 10 watts |
| Volume | 7 by 4.7 by 2.4 inches (196 x 120 x 66 millimeters) |
| Data Return | 5 to 10 kilobytes per sounding location |
| Frequency Range | 150 to 1200 megahertz |
| Vertical Resolution | As small as about 3 to 12 inches thick (15 to 30 centimeters) thick |
| Penetration Depth | Greater than 30 feet (10 meters) deep depending on materials |
| Measurement Interval | About every 4 inches (10 centimeters) along the rover track |

Table 2: RIMFAX Ground Penetrating Radar System Characteristics. Reprinted from [18].

| Location | Mounted on the rover mast at the eye level of a 6 ½-foot-tall person (2 meters tall). The cameras are separated by 9.5 inches (24.2 centimeters) to provide stereo vision. |
|---|---|
| Mass | Approximately 8.8 pounds (about 4 kilograms) |
| Power | Approximately 17.4 watts |
| Volume | Camera head, per unit: 4.3 by 4.7 by 10.2 inches (11 by 12 by 26 centimeters) Digital electronics assembly: 8.6 by 4.7 by 1.9 inches (22 by 12 by 5 centimeters) Calibration target: 3.9 by 3.9 by 2.7 inches (10 by 10 by 7 centimeters) |
| Data Return | Approximately 148 megabits per sol, average |
| Color Quality | Similar to that of a consumer digital camera (2-megapixel) |
| Image Size | 1600 by 1200 pixels maximum |
| Image Resolution | Able to resolve between about 150 microns per pixel (0.15 millimeter or 0.0059 inch) to 7.4 millimeters (0.3 inches) per pixel depending on distance |

Table 3: Mastcam-Z Camera System Characteristics. Reprinted from [17].

**Exploration Preparation**

Lava tubes have so far only been explored from above the surface by mapping the cave entrances using data from Martian orbiters, as briefly mentioned previously. As a result, mapping and exploring a lava tube with a rover mission would be the next step in gaining a better understanding of these important structures. Drawing on the orbiter's location data, characteristics, such as the shape of a lava tube's entrance, could subsequently form the basis for decisions concerning further exploration using micro-rovers stored as a payload in the main Martian rover.

The rover will be able to drive to the entrance from a nearby landing point. The Mastcam-Z camera system attached to the main rover can send images of the tube entrances to Earth, where experts can assess the cave's exploration value. Large caves with an entrance and structure that allow for human exploration and habitation, for example, would constitute an ideal mission location.

Furthermore, the RIMFAX ground penetrating radar system could estimate the size and depth (up to 10 - 20m, [19], [18]) of the tube from above the surface, assisting experts in their decision-making process.

After identifying and selecting the lava tube's entrance through the inspection completed by the main rover, a detailed inspection of lava tubes could follow through our rover mission, including the deployment of small robot swarms inside the lava tube.

**Scientific Return**

Even though both systems have already been utilized for the Perseverance rover mission, the Mastcam-Z and RIMFAX can both yield important additional insights due to the distinct landing location of our rover and the unique environment inside the lava tubes.

Overall, the Mastcam-Z camera system yields important additional visual information about the Martian environment. Camera data can likewise be used to conduct a spectral analysis of various rock samples on the surface of Mars to aid mineral identification. Here, [20] and [16] suggested that camera data collected by the proposed camera system can successfully be used to estimate different minerals. Furthermore, camera data can, for instance, be used to monitor the terrain and atmosphere of the Martian environment (See table 4 for a detailed description of various camera objectives, please refer to [16].)

| Mastcam-Z goals | Mastcam-Z detailed investigation objectives |
| --- | --- |
| 1. Characterize the overall landscape geomorphology, processes, and the nature of the geologic record (mineralogy, texture, structure, stratigraphy) at the rover field site | 1-a. Characterize the morphology, texture, and multispectral properties of rocks and outcrops to assess emplacement history, variability of composition, and physical properties. |
| | 1-b. Determine the structure and orientation of stratigraphic boundaries, layers, and other key morphologic features to investigate emplacement and modification history. |
| | 1-c. Characterize the position, size, morphology, texture, and multispectral properties of rocks and fines to constrain provenance and weathering history. |
| | 1-d. Observe and monitor terrains disturbed by rover wheels and other hardware elements to assess surface to physical and chemical weathering. |
| | 1-e. Distinguish among bedform types within the vicinity of the rover to evaluate the modification history of the landscape. |
| | 1-f. Identify diagnostic sedimentary structures to determine emplacement history. |
| | 1-g. Characterize finer scale color/spectral variation (e.g., cm-scale veins, post-depositional concretions) to constrain provenance and diagenetic history. |
| 2. Assess current atmospheric and astronomical conditions, events, and surface-atmosphere interactions and processes | 2-a. Observe the Sun for rover navigation and atmospheric science purposes. |
| | 2-b. Observe the sky and surface/atmosphere boundary layer to measure atmospheric aerosol/cloud properties and transient atmospheric/astronomical events. |
| 3. Provide operational support and scientific context for rover navigation, contact science, sample selection, extraction, caching, and other Mars 2020 investigations | 3-a. Acquire stereo images for navigation, instrument deployment, and other operational purposes on a tactical timescale. |
| | 3-b. Acquire sub-mm/pixel scale images of targets close to the rover. |
| | 3-c. Resolve morphology and color/multispectral properties of distant geologic features and topography for longer-term science and localization/navigation planning purposes. |

Table 2: Mastcam- Z Objectives. Reprinted from [16].

The ground-penetrating radar system could additionally be used to identify water resources on Mars. [21] and [22] proposed that using ground-penetrating radar to map such structures on Mars could be as successful as it is on Earth. The radar system in combination with camera data could also be used later in the mission to investigate potential water-ice sources [19] and could yield important insights into the ancient development of the Martian surface [18].

**Costs**

As the development of the Mastcam-Z and the RIMFAX system are already completed, only adjustments to the hardware have to be made. No information about the specific costs of the Mastcam-Z system and the RIMFAX system appears to be available. However, information about sub-components can be found or inferred. E.g., the optical sensor used in the Mastcam-Z KAI-2020 by Kodak was listed by vendors for about $3500. Prices for the other components cannot be found directly and can therefore only be broadly estimated. The RIMFAX system is separated into two modules, the processor BAE RAD6000 which costs about $300.000, and the ground low-gain x-band antennae manufactured by enduroSAT. No prices were released by enduroSAT, nevertheless prices for ground penetrating sensors can go up to $50.000. Please refer to table 5 for a detailed estimation of costs for the camera system, and to table 6 for the RIMFAX system based on information found online.

| Component | KAI-2020 optical sensor (Kodak) | Lenses | Other components | Development & Integration | Total |
|---|---|---|---|---|---|
| Cost estimate ($) | $3500 | $50.000 | $40.000 | $100.000 | **$193.500** |

Table 3: Cost Estimate ($): Mastcam-Z.

| Component | BAE RAD6000 | Ground Sensor | Other components | Development & Integration | Total |
|---|---|---|---|---|---|
| Cost estimate ($) | $300.000 | $50.000 | $50.000 | $100.000 | **$500.000** |

Table 4: Cost Estimate ($): RIMFAX.

**Schedule**

As the Mastcam-Z and the RIMFAX ground penetrating radar system have already been planned to be used in the Mars 2020 missions, it is ensured that the goals of the missions as well as the development process of the technologies will be completed before 2033.

**PAYLOAD 2: LAVA TUBE EXPLORATION ROBOTS**

**Engineering Design**

Prior research into potential Mars exploratory vehicles has presented various robot designs, ranging from flying drones [23] to autonomous mini-rovers [24] to hopping or rolling robots [25], [26]. However, in the context of Martian lava tube exploration, the German Aerospace Center (DLR) proposed an extremely robust robot designed specifically for the challenging environment (such as debris inside the tube) [27]. The snake-like robot is made up of various modules, each with rimless wheels attached to the side. Each robot is made up of a single main module that houses a power and communication source.

The side modules have their own actuators and can carry up to 6kg of payload. A flexible connector between the various modules provides shock absorption and flexibility, allowing the robot to be dropped inside the tube from a height of approximately 1.5m. If one of the rimless wheels becomes stuck inside the debris, it can be automatically removed from the robot's body and the exploration mission can continue as planned [27].

To collect data from inside the tube, the robots can be equipped with cameras and spectroscopic technology. Furthermore, various mapping technologies could be used to map the inside of the tube, such as LIDAR systems. However, using the space inside the rover's side elements, it could be outfitted with other tools for analyzing minerals or collecting samples. The samples could then be transported back to the cave's entrance and analyzed inside a station lowered inside the cave and containing the gas chromatograph instrument (see section 7), which could also serve as a charging and communication station for the rovers (see gravitational battery station system). Multiple robots could be lowered inside the tube to map the environment efficiently and without overlap while communicating using a SLAM-based algorithm ([28].

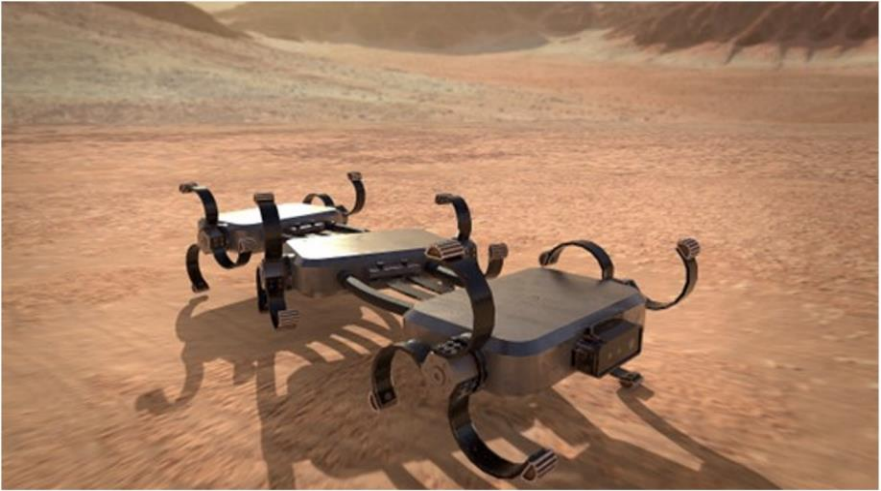

Figure 1: DLR Lava Tube Scout Robot based on [27].

| | |
|---|---|
| Rover mass | $\sim$18$kg$ |
| Payload capacity per aux module | 6kg, 51 volume |
| Max. speed | $\sim$1.7 $m/s$ |
| Max. obstacle height | >400$mm$ |
| Tested drop height | >1.5$m$ |
| Battery run time (single/ double) | >5h/10h |
| Number of modules | 2-5 |

Table 5: Scout Rover Characteristics. Reprinted from [27].

## Exploration Preparation

The topology of the lava tubes raises as a key interrogation if they are considered as potential anti-radiation shelters for astronauts living on Mars. Mapping their interior through LIDAR sensors or understanding their geological nature is a critical prerequisite for initiating any habitat design activities or focusing on other solutions with respect to astronaut protection. Such data could even be used to recreate the lava tubes in immersive virtual reality to facilitate training of future astronauts going to Mars.

## Scientific Return

The scouting rover shall be able to grab geological samples deep in the lava-tube and bring them back to an analyzer. Such samples would, in turn, facilitate greater understanding of the geological nature and history of this volcanic region. Additionally, this procedure might enable us to analyze possible water samples and minerals located inside the lava tube. As each element of the robot can contain up to 6kg of payload, each robot can collect several separate samples inside the structure without contamination risks between samples. Therefore, the robot design allows for maximum science return while avoiding unnecessary risks by minimizing travel distance.

## Costs

As the costs for the development and the manufacturing of the DLR scout rover has not been made available to the public, no definite price for the system can be determined. However, as the development of the robotic platform is already in an advanced stage, we can assume that the costs for further development will be comparatively low.

Costs for NASA's Mars rovers, on the other hand, have been made public. Based on the DLR scout rover's complexity and size, we can deduce that the development costs of NASA's Pathfinder rover of $174,2 million would be roughly comparable.

## Schedule

Importantly, the proposed rim-less scout rover concept is already in advanced stage of development, with the concept scheduled to be mission ready in 2030. The ARCHES analog mission, a collaboration between DLR and ESA, will conduct initial tests this year inside a lava tube on Earth [27].

**PAYLOAD 3: WIND POWER BALLOON**

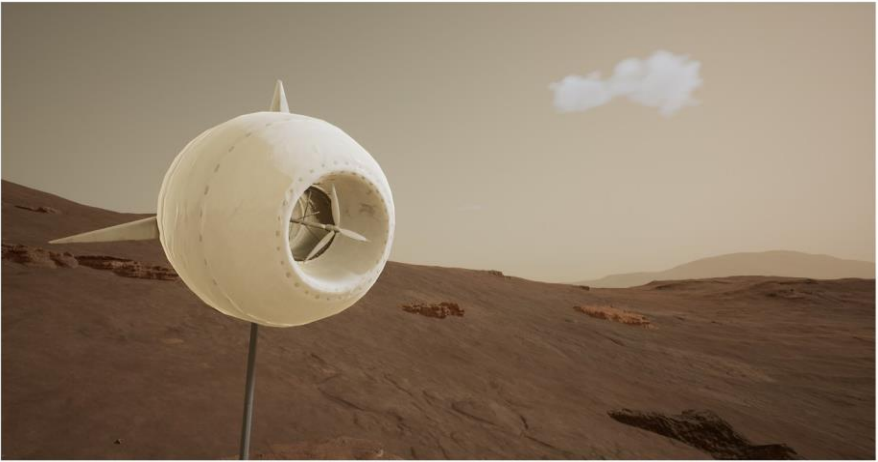

Figure 2: Wind Power Balloon Illustration.

The underlying concept here is very simple: A wind turbine similar in shape to a turbine of a passenger jet-plane but with a vastly bigger size, where the tube-shaped outside is an inflatable balloon. Alongside the turbine, multiple scientific payloads, provided the buoyancy permits it, can also be mounted onto it.

**Engineering Design**
The balloon will be anchored to the ground by an extensible power cord, allowing for the adjustment of altitude of the balloon. Mounted inside the tube sits the turbine, with rotors and the generator, with a significantly lower weight compared to a similar terrestrial turbine, due to the atmospheric density being substantially lower. Next to the turbine it is also conceivable to attach various scientific instruments to monitor atmospheric composition, pressure, temperature, and particles, such that are being flung around during Martian storms. Finally, such a platform could wear a coating made of electrodes to allow electrons from the ground to flow through the material of the tether, due to the difference of electric potential between the ground and the relatively high potential of the Martian atmosphere, thus creating an electric current.

The characteristics of the Martian atmosphere provide unique challenges and opportunities compared to flying a balloon on Earth. The biggest challenge is by far the extremely low density, making it more challenging to achieve buoyancy, when carrying heavy equipment, such as the turbine or a scientific payload, requiring the lifting volume to be scaled up. On the flip side, the low density also equals a lower force of wind, where a storm on Mars is comparable to a summer's breeze, lowering the danger of damage even in storms and allowing for lighter materials in the anchoring, hull of the balloon and turbine but also lowering the expected power output. The temperature is also a notable difference, ranging from about 20C° to around -100C° at its coldest around equatorial regions. This should be kept in mind when selecting equipment, material and lifting gas.

Speaking of lifting gasses: One major advantage of the Martian atmosphere over the Terran one is that it consists almost exclusively out of $CO_2$, having a comparatively high molar mass of around 44g/mol. Any gas with a lower molar mass is a potential lifting gas, including the popular Helium (4g/mol), but also Oxygen ($O_2$ has around 32g/mol) and Hydrogen. A problem with Oxygen, however, is that it can be very reactive and corrode the hull over time, particularly when being transformed into Ozone by the very present radiation. Perhaps the biggest advantage of Oxygen is the in-situ availability, where it can be generated by either the greenhouse payload (see payload 6) or the already demonstrated MOXIE payload that could also be mounted on a big version of the balloon to autonomously regenerate its lifting gas anywhere. With an anchoring mechanism and a control circuit using the turbine for locomotion, such a balloon may serve as a permanently airborne independent drone, as proposed by [29].

Hydrogen is also a good candidate, which can be found on Mars in small quantities and is not flammable in the Martian atmosphere but has the downside of easily escaping the lifting volume, due its elusiveness. Helium, whereas it has very favorable characteristics in reactivity, elusiveness and lifting capabilities, it will have to be brought entirely from Earth, as the Martian atmosphere is devoid of it in low altitudes. The balloon covering may be selected according to the lifting gas, but a black coloration for the absorption of radiation may be favorable, also in increasing the lift during the day due to the heating of the lifting gas by the increased absorption of sunlight. Table 8 lists a possible configuration of such a setup, with the assumption of a perfect cylindrical shape for the lifting volume, which ideally should be aerodynamic to facilitate better wind-flow and using $O_2$ as the lifting gas. It should be noted that the final values are expected to vary based on selected payloads and lifting gas.

The overall density in this example is lower than the Martian atmospheric density at about 0,02kg/m$^3$, resulting in a positive buoyancy.

A detachable design has also been considered with an anchor, that can grapple the base station or attach itself to Martian rocks or even into the soil. Since the balloon has its own power source and a built-in turbine, it is conceivable for it to be an independent drone with a flight range and duration tied to its capability to retain or regain its own lifting gas. This however would make the drone more complex especially if it is supposed to have controls and avionics. The feasibility at this stage will have to be investigated further and may prove too challenging due to its engineering requirements.

| Variables | Value |
|---|---|
| Outer radius $(m)$ | 7 |
| Inner radius $(m)$ | 3 |
| Length of tube $(m)$ | 6 |
| Lifting gas density $(kg/m^3)$ | 0.008008584 |
| Surface area weight $(kg/m^2)$ | 0.01 |
| Tether length (m) | 40 |
| Tether weight per length $(kg/m)$ | 0.01 |
| Scientific payload weight $(kg)$ | 2 |
| Windmill weight $(kg)$ | 2 |

Table 6: Wind Power Balloon Configuration Set-Up.

**Exploration Preparation**

The turbine and atmospheric electricity-based generator will provide another option of power generation for Martian infrastructure as well as an option to monitor the environment from an intermediate altitude, allowing for payloads requiring a higher proximity for their measurements. It also provides a test environment for buoyancy-based aviation that can be controlled and tested over longer periods of time from the base itself. Both the power generation by wind and transportation aspects have been around for a long time on Earth but have never been tested on Mars. This mission not only makes both possible at once, but also makes wind-power more feasible by requiring minimal supporting infrastructure for variable altitudes and a way to funnel the very low intensity Martian winds into the turbine, while exploiting an alternative power source in the form of atmospheric electricity simultaneously.

**Scientific Return**

The primary scientific return of the balloon will stem from its monitoring and observations of the atmosphere. While it may rest at high altitudes by default, ultimately the balloon can be put at desired altitudes for different tests. Results, especially during sandstorms, might be of particular interest. Moreover, in terrestrial environments, it is feasible for a small corona motor to be powered by exploiting the electric potential of the atmosphere. It could then be rewarding to verify if the same principle would work on Mars under fair-weather conditions. Additionally, the balloon may be employed to assist with scouting and locating suitable lava tube entrances, beginning shortly after deployment on Mars.

**Costs**

The cost is difficult to gauge due to variabilities in the possible setup. In the following table an estimate of costs for the most complex configuration is proposed. Since this subsystem is still in its early stages of development, it can be assumed that salaries will be the biggest expense during the development and integration phases over several years. It should be noted here that most of the cost depends on the scientific payloads used.

| Element | Lifting gas (O2 tank) | Surface covering | Tether system | Turbine system | Scientific payloads | Development | Integration | Total |
|---------|----------------------|------------------|---------------|----------------|---------------------|-------------|-------------|-------|
| Cost estimate ($) | 350 | 1000 | 500 | 500 | 5000 | 2.000.000 | 100.000 | 2.107.350 |

Table 7: Estimated Costs - Wind Power Balloon System.

**Schedule**

While it is not easy to pinpoint a schedule based on previous missions or similar experiments, the principles and engineering are very simple and most parts, except for the custom-made hull and an adjusted inflation-system, can be bought off the shelf with minimal adjustments. Since it also does not need to be pressurized during transport, the storage-equipment can also be very minimal. Therefore, it can be argued that a launch by 2033 is very reasonable as the required development should be rather minimal.

## PAYLOAD 4: MYCOTECTURE

The idea of mycotecture (a combination of the words *Mycelium*, the "root" part of fungi, and *architecture*) represents a relatively novel approach to producing building material and for building, or rather growing, structures. Here the properties of fungi to "mold" into any shape, filling it completely with fungal mass that is sturdy but also relatively light weight while only requiring oxygen, water, and nutrients, is being used to create partial or entire structures. This concept has been investigated by both NASA [30] and ESA [31] among others [32].

**Engineering Design**

To provide the growth environment, the following resources need to be available: A mold or inflatable structure for the mushroom to inhabit, water, nutrients, oxygen, and the right thermal conditions.

The mold can come in two different forms: A rigid block to create one or multiple uniform bricks of mycelium or an inflatable structure that can then be filled out by the mycelium. A hybridization of both is also thinkable.

Water, nutrients, and oxygen can come pre-processed from earth as part of the setup or be possibly sourced from the greenhouse payload (see section 7). Whereas the latter option forms a nice synergy with the greenhouse payload, the former option should always have preference for the sake of redundancy. Oxygen may successfully be sourced from the greenhouse payload (see payload 6), should its deployment reach a significant level of success.

Thermal control might constitute the biggest challenge, one that is not widely discussed in existing literature, which seems to generally presuppose optimal temperature. One obvious requirement is for the growing environment not to freeze over. How low the temperatures can go without damaging the fungi is however difficult to assess. On the other hand, low temperature can be exploited for "deactivating" the mycelium simply by exposing it to the Martian climate without providing additional heating.

A potential mycotecture system could have these estimated characteristics during transport:

| Mass | Volume | Energy consumption |
|------|--------|--------------------|
| 50kg | <1m^3 | ~3W, if insulated well |

Table 8: Mycotecture Payload Engineering Design.

## Exploration Preparation

The notion of growing fungi in off-world conditions has seen a surging interest in recent years, with relevant studies ongoing. Our overall aspiration with this technology is for it to provide shelter and accommodations for future Martian inhabitants. The risk of these attempts failing, however, are considerable. Therefore, it is crucial to assess mycotecture solutions thoroughly, proving that they are viable on Mars before they are relied on during subsequent crewed missions. In addition, fungi are known to be able to extract minerals from soil, which will also be crucial for foraging vital resources for many purposes of Mars colonization. In the light of the great potential and the great unknowns still surrounding this technology, we can safely say that there is a strong incentive to studying it further.

## Scientific Return

The main scientific interest with fungi in the context of this mission will be to examine techniques under which fungi, specifically mycelium, can grow. This includes figuring out ideal conditions on Mars for growth as well as trying to use in-situ resources, such as soil, oxygen from photosynthesized or electrolyzed carbon-dioxide and possibly water from methane and oxygen, if available. This all links to the very central question of how life can thrive on Mars.

## Costs

Since there is no suggestion on a concrete design for a mycelium growing system for Mars, this estimate tries to give a rough idea with generous margins included in the calculation. To try as many different techniques as possible, both an inflatable system and a solid one will be covered. The mycotecture will use a similar setup as can be found in the greenhouse payload. This is to reduce development cost, as it can be reasonably assumed that the mycelium thrives in a similar, if not the same, environment, as the greenhouse plant. Due to the early development stage of this subsystem, the biggest costs are expected in HR for development and integration over several years.

| Element | Grow volume (solid and inflatable) | Resource supplementation syst. | Nutrients | Temp. control syst. | Redundancy | Development | Integration | Total |
|---------|-----------------------------------|-------------------------------|-----------|---------------------|------------|-------------|-------------|-------|
| Cost estimate ($) | 1500 | 1500 | 500 | 100 | 500 | 2.000.000 | 100.000 | 2.104.100 |

Table 9: Costs Estimates for the Mycotecture Payload.

**Schedule**
The technology behind mycotecture may be young but has been, and currently is, thoroughly investigated, having even been trialed on the ISS. As of today, no trial has however taken place on Mars. This means that there still is some development work required, but a launch by 2033 is still realistic, considering the current efforts.

## PAYLOAD 5: GAS CHROMATOGRAPH

**Engineering Design**
As previously mentioned, lava tubes are more likely to host life or at least hold some traces of past life on the Red Planet. The Curiosity rover was looking for traces of life using the SAM (Sample Analysis at Mars) tool and our team plans to add the same kind of instrument to make sure our rover can track life forms in Lava tubes. Therefore, a gas chromatograph will be added to analyze the atmosphere inside the lava tubes. This technology does not present a substantial engineering challenge as it has already been designed and sent to Mars with Curiosity [33]. The characteristics of such an instrument may be as follows: A mass of 35kg, dimensions of 46cm x 27cm x 29 cm and a required power of 120W.

**Exploration Preparation**
Analyzing the inner atmosphere of lava tubes may be critical to the explorations task of the astronauts. Indeed, this may help to spot dangerous gasses that would appear in high concentration, in particular explosive ones.

**Scientific Return**
One of the gasses that is released by lifeforms is methane and for now only small concentrations of it have been detected in the Martian atmosphere by the rover Curiosity [34]. Potentially significant concentrations of methane have also been detected over the Nili Fossae region by Earth-based telescopes [35][36]. The origin of this methane may also be geological [37], and our mission may answer the question of its origin by bringing the Gas Chromatograph to Mars again, and more specifically to Nili Fossae.

**Costs**
Concerning the price of such an instrument, we will rely on the technology developed for the Curiosity rover. The provided costs estimate would then only include the acquisition of the components. Based on current market prices, we estimate this would translate into approximately 50,000$.

**Schedule**
Finally, this technology is already available and ready for deployment. Only the integration time and adaptation of the instrument to get samples from the lava tube exploration robots (see section 4) may require more time. However, those tasks would only require a few years, mostly for the integration and to create the anchor point for the robots. Such a development is reasonably achievable for a launch by 2033.

## PAYLOAD 6: DEPLOYABLE GREENHOUSE

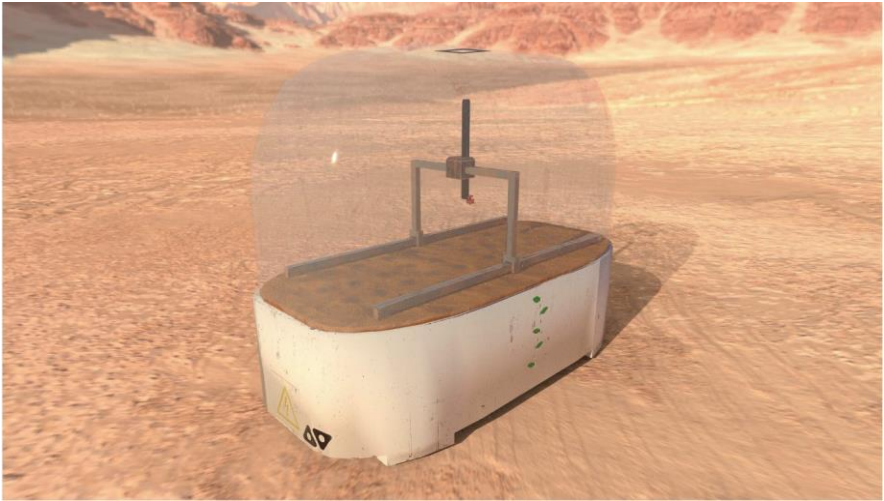

Figure 3: Concept for the deployable greenhouse

### Engineering Design

The next payload that we are planning to bring to Mars is a Greenhouse that would be deployed on the surface to test full soil growth of plants. This experiment has a notable synergy with the mycotecture payload, as it will be able to supply the mycelium (payload 4) with oxygen which is critical for their growth. Two main elements are required for this payload: first, an automated greenhouse that can deploy and plant seeds in the Martial soil, and subsequently support the growing plants. The second element is a set of seeds from species whose growth would already have been tested into Mars soil simulants to give them the best chances to survive and grow.

Regarding the hardware setup, our design will mostly rely on what the open-source community came up with in the context of the Farmbot initiative [38]. It consists of a robotic autonomous farm that has a growth area as big as $18m^2$. Through its raspberry computer, it can monitor the growth of the plants while providing them with everything they need. The possibilities are even wider as some projects use the Farmbot in association with photogrammetry technologies to have a 1-to-1 digital twin of the garden. What we intend to add to this setup is a life support system (temperature control, atmosphere monitoring …) to allow the greenhouse to survive Martian nights, but also some redundancy in case of system failures.

An area of 2x1m is conceivable to give enough space to the plants. The structure would then be 1.5m tall. The payload fits in a volume of 1x2x1.5m or 3 cubic meters. Once deployed on soft ground, the bottom part of the structure could start collecting the Martian soil to put it inside the greenhouse thanks to a bucket-wheel excavator. This technology appears to be a reliable choice for such a mission, as presented by the trade-off study realized by García de Quevedo Suero et al in the context of a Martian greenhouse-rover design project [39]. The farm bot would then
spread the dirt on the whole surface, using a rake tool, from the movable platform
that brought it in the grow volume (see figure 4).

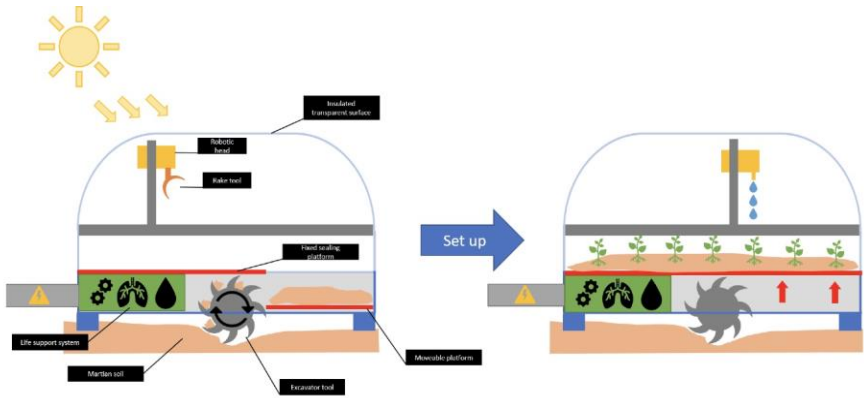

Figure 4: Greenhouse Design Concept, during (Left) and after (Right) Set-Up.

The Farmbot's energy consumption is estimated to be 0.287kWh/day by its
manufacturer. In addition, to handle the temperatures that can drop to -73°C during
the night, the life support system should be able to maintain the temperature around
20°C. The design results are reported in the table 12. Regarding the power required
to maintain the temperature in the greenhouse, extensive efforts may have to be
done on the dome insulation, using insulated glazing either by Mars atmosphere or
by vacuum. These technologies would provide even lower heat transfer coefficients
than the one that has been used for the design here (air insulated glazing).

| Glassed Area | Double glazing heat transfer coeff. | Target temperature | Night temperature | Estimated heat transfer | Night Duration | Required energy /night |
|---|---|---|---|---|---|---|
| ∼5m² | ∼1.1 W/m².K | 20°c | -73°c | 511.5 W | 12h20mins | 6.3 kW.h |

Table 10: Greenhouse Engineering Design.

The energy consumption of the excavator has not been addressed here because all
other systems would be shut down during the excavation process and following the
current design it will have enough power to run. Looking at the species to bring,
studies have shown that nitrogen-fixer species, such as Lotus Pedunculatus, would
enrich the soil with this critical nutrient for other species to grow. Moreover, some
Crops species like Lepidium sativum are good candidates to grow on Martian soil.
Both showed germination rates above 70% in Martian soil simulants [40]. Finally,
the mass of the total setup may not go over 150kg, with approximately 30kg
reserved for the Farmbot and 120kg for the structure and life support system.

| Element | Farmbot | Deployable feature | Glass | Temp control syst. | Redundancy | Development | Integration | Total |
|---|---|---|---|---|---|---|---|---|
| Cost estimate ($) | 3.500 | 2.000 | 500 | 100 | 500 | 2.000.000 | 100.000 | 2.106.600 |

Table 11: Cost Estimate for the Greenhouse.

**Exploration Preparation**

This experiment would assess the possibility to grow plants in Martian soil, which will become critical to foster human exploration and colonization of this planet. If it works, it could also provide the other payloads of the mission with oxygen if required.

**Scientific Return**

This will be the first time that organisms from Earth grow on Mars. We may furthermore identify those species that are the most suitable for growing on Mars by bringing seeds from different plants that have been demonstrated to grow in Martian regolith simulants. However, the growth in regolith may not go as smoothly as expected [41] and failure would also give valuable data contributing to greater chances of success during future attempts. This is a unique chance for a full soil plant growing attempt. On the other hand, this experiment introduces a high risk of Martian soil contamination in case of failure of the greenhouse. Thus, planetary protection protocols must be followed rigorously to avoid any contamination.

**Costs**

Concerning the cost of such a set-up, Farmbot-wise, the $4.5m^2$ version of the robot would cost around 3500$. However, our design would make the structural elements of the set-up deployable (e.g., using a dedicated crane) which represents an additional cost estimated at around another two thousand dollars, even if we plan to have a smaller growth area. There's also a need for an environmental control system and some redundancy. Since the system is in an early stage, the biggest costs could be expected for development and integration. The cost breakdown is reported in table 13.

**Schedule**

About the technologies' readiness, the Farmbot products should be available no later than summer 2022, and all other elements constituting the design we propose are already available on the market at reasonable costs. Once the Farmbot is received, we estimate that 4 years would be required to make the system Mars ready by adding all the aforementioned features. This leaves even enough time to try out this greenhouse design on the Moon first before sending it to Mars.

**PLATFORM/SYSTEM (ROVER(S)/BASE(S))**

**Winch System and Gravity-Based Power Generation**

This section deals with the engineering design of the winch system. This tool will allow our rover to go down into the lava-tube to set up the exploration robot swarm's station and then to go back up again. This system would be able to take advantage of the gravitational potential energy turned into kinetic energy during descent, thus generating power as it travels down the tube's walls.

In terms of technology readiness, winch systems are already mature and commonly used on Earth. Their main components are a direct current motor, an electronic board to control it, some gear systems for power transmission, a shaft and a cable/rope wound around it.

The "worst-case" scenario we considered while designing this system was the rover running into problems while climbing during the descent. The rover should then be able to bring the payload back to the central station tethered at the top of the lava tube. The application of Newton's 2nd law to the payload-winch system leads to the results presented in table 14 below:

| Payload mass | Ascend/Descent speed | Vertical distance | Required winch motor power (+10% margin) | DC motor round trip efficiency as generator | Generated energy per descent |
|---|---|---|---|---|---|
| 500 kg | 40 cm/s | 100 m | 0.746 (0.820) kW | 70% | 36,24 Wh |

Table 12: Winch System Requirements and Power Generation Capabilities.

According to Boretti et al [42], in 2013 the round-trip efficiency, meaning from wheels to battery and wheels again, of a regenerative braking like the one we would like to use was estimated at most, around 70% for electric cars. Note that dedicated dynamo can have efficiencies up to 80%. It must also be considered that this estimation is highly dependent on how people drive the cars. In our case, however, we can assume that the rover goes down in a way that maximizes energy generation. This additional energy generation system won't clearly be enough to provide other systems with power, nonetheless it covers some of the power requirements of the winch system. Moreover, winch systems fitting those requirements may be found on the market with the characteristics reported in table 15.

| Model | Price | Mass | Electric consumption | Dimensions (LxWxH) |
|---|---|---|---|---|
| L-GT300-SY-12V | ~400$ | 17kg | 1700kW | 391mm x 126mm x 128mm |

Table 13: Example of an off the Shelf Winch System Fitting our Requirements [43].

By adding additional redundancy (another motor) and the possibility to transmit data and power through the winch cable, as part of the central station energy generation endeavor, it must be feasible to design a similar system in terms of dimensions and electric consumption at an additional cost estimated around 300$.

**Avionics and Sensors**
The avionics unit will contain all the electronics of the rover that handles data and resources transmission between the different payloads. It will process information from the sensing organs of the rover that will be added: the 360° camera, thermometers, pressure sensors, and microphones. This unit is also responsible for communication with Earth through an antenna. Particular attention may then be paid to the integration and physical protection of this part, this is the reason why it is planned to have it at the center of the main station deployed at the entrance of the lava tube. Furthermore, this is a strategic location in the sense that it will then be directly connected to all payloads. Embedded systems and their protection may weigh around one metric ton according to current rover designs (Perseverance [44]). The avionics unit is not able to handle temperatures out of the range -40°c/+40℃ on the contemporary generation of rovers [45]. On Mars, the temperature may drop to -90°C during nights which would cause the rover to freeze to death without the proper equipment. Indeed, the electronics will release some heat over time, thus making use of gold paint, good insulation and heaters will help maintain the temperature of the components in the acceptable range.

## Chassis and Navigation system

Concerning the navigation system, the proposed design does not reinvent the wheel: It will be equipped with a chassis similar to the current generation of Martian rovers with 6 motorized wheels, also known as the rocker-bogie design, that will allow it to move on the chaotic terrain of the lava tubes especially. The chassis will be modular, allowing the payloads to be deployed individually onto the Martian surface. This element will concentrate most of the rover's mass. It must also be mentioned that the chassis will have to be big enough to host all payloads in it, which corresponds to approximately 270kg and a volume close to 5 cubic meters.

## Power Generation

This part focuses on the elements that will generate and store electricity. Some of the systems that will be carried can generate power: The winch system and the wind balloon. However, these are not producing enough power to supply all other systems in the mission. The elements that would require constant energy supply to work are the rover's "brains" (avionics), the greenhouse, the mycotecture and the drone control station. Thus, they are the one critical to the power supply design. The other components only need energy at one point, but we still need to make sure the rover's power generation system would provide them with enough power at the right time.

A Radioisotope Thermoelectric Generator (RTG) can produce a few hundred watts, like 110W for the Perseverance rover, which would be sufficient to cover our mission's power requirements. On Perseverance this unit weighed 45kg for 64cm diameter and 66cm length. This represents a source of energy that is independent of current lighting conditions, unlike solar panels. One of the drawbacks of that technology is that it releases heat, but this can be turned into an advantage if that wasted heat, which is not turned into electricity, could be used for thermal regulation purposes, particularly during nights.

To conclude on the platform design, it is hard to give an estimate on how much the chassis will weigh, but at this stage of our mission mass estimate, more than 9 tons are available to the platform design with respect to the lander capabilities, which should comfortably cover the needs of the chassis.

**SCHEDULE: PROJECT LIFE-CYCLE**

This project is now in phase "Pre-A" as per the NASA Procedural Requirements NPR 7120.5F [14] "Subject: NASA Space Flight Program and Project Management Requirements". The NASA project life cycle is comparable to the waterfall methodology [46]. Key Decision Points A (KDP A) must be established and approved by the project's decision authority to move on to phase "A" of the project. No phase will be repeated, in contrast to agile approaches like SCRUM [47].

This project's experiment payloads are the main topic of this work. Given the estimated schedule we provided for each payload, it has been determined that launching the mission by 2033 should be feasible. Based on historical data (e.g., schedule of Perseverance rover), it can be assumed that other processes, such as implementing flight safety standards in accordance with NPR 7120.5F, shouldn't cause schedule delays. A GANTT chart in table 16 shows an example timeline for all planned phases A through F with a launch at the end of phase D in 2033.

| Phase | Life-Cycle Phases \ Year (20XX) | '22 | '23 | '24 | '25 | '26 | '27 | '28 | '29 | '30 | '31 | '32 | '33 | '34 | '35 | '36 |
|-------|--------------------------------|-----|-----|-----|-----|-----|-----|-----|-----|-----|-----|-----|-----|-----|-----|-----|
| Pre A | Concept Studies | | | | | | | | | | | | | | | |
| A | Concept Technology Development | | | | | | | | | | | | | | | |
| B | Preliminary Design | | | | | | | | | | | | | | | |
| C | Final Design Fabrication | | | | | | | | | | | | | | | |
| D | Assembly, Integration & Verification | | | | | | | | | | | | | | | |
| E | Operation Sustainment | | | | | | | | | | | | | | | |
| F | Closeout | | | | | | | | | | | | | | | |

Table 14: Project Lifecycle: GANTT Chart.

The primary mission (phase E) will have the following structure depending on the intended payloads:

1. Initial phase: The rover system will land on Mars on a relatively safe spot, which is in reasonably proximity of probable locations of lava tube entrances. After landing, routine self-assessment and tests will be commenced in conjunction with the deployment of mobile means of power generation and reconnaissance. This includes solar panels, the wind-turbine balloon and lava tube exploration robot swarms.
2. Transit phase: Next the rover will have to relocate to a lava-tube. It will do so by driving there, surveying the environment. Navigation is intended to be done by GNSS, if available at that time, or similar means, with a preselected location based on already available observations of Martian geography.
3. Settlement Phase: This phase encompasses two major components: The setup and usage of the greenhouse and mycotecture payloads as well as the lava tube exploration. This is done by setting up the base station at the entrance of the lava tube and by deploying the greenhouse along with the mycotecture environment. From there the winched rover will travel inside the lava-tube carrying the exploration robot swarm. These will subsequently map and survey the lava-tube.

By remobilization of the greenhouse and mycotecture payload, the Transit and Settlement phase can be looped to explore multiple lava-tubes in the area.

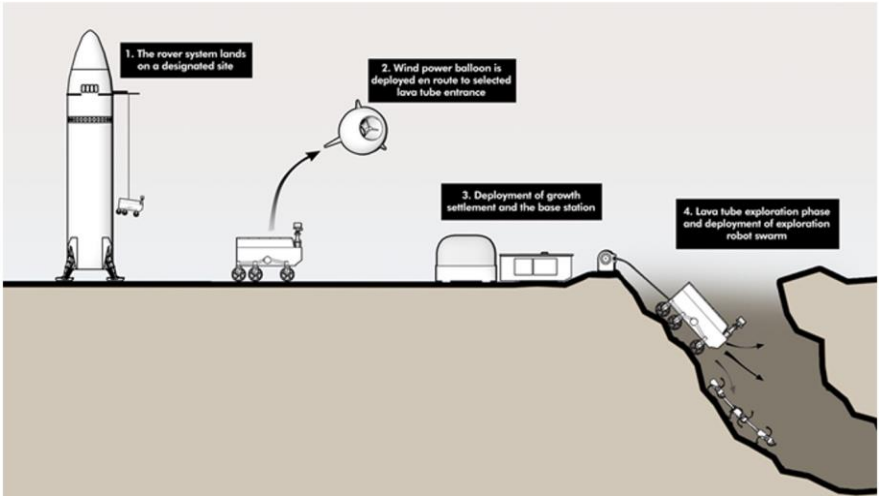

Figure 5: Schematic Mission Plan Illustration.

## PROJECT LIFE-CYCLE COST ESTIMATE (LCCE)

We have provided cost breakdowns for several components. However, overall costs are only first estimates. These estimates are based on data from previous missions, or vendor data. As defined by the challenge, the LCCE is limited to the costs generated by the payloads. An LCCE for the payloads up to Work Breakdown Structure (WBS) level 3 can be seen in table 17.

For external stakeholders, complete time-phased cost plans and schedule range estimates up to WBS level 2 [15], should be ready at Key Decision Point phase B (KDP-B) and high-confidence cost and schedule commitments at KDP-C according to NPR 7120.5F [14]. These estimates in phase B and C of the NASA project life cycle can be calculated using SEER-H by Galorath [48] or, alternatively, cost estimates could be created based on the NAFCOM (NASA Air Force Cost Model) [49].

| Name | Ground Penetrating Radar Camera | Lava Tube Exploration Robot | Wind Power Balloon | Mycotecture | Gas Chromato-graph | Deployable Greenhouse | Payloads |
|---|---|---|---|---|---|---|---|
| **WBS Level** | Level 3 | Level 3 | Level 3 | Level 3 | Level 3 | Level 3 | Level 2 |
| **Cost estimate** | $693.500 | $174.200.000 | $2.107.350,00 | $2.104.100 | $205.453 | $2.106.600 | $181.417.003 |

Table 15: LCCE for payloads up to WBS level 3.

**CONCLUSION**

It can be argued that all payloads discussed here are well within given requirements of volume and mass, even with a generous margin of error added onto them, with room to spare for other missions. The total required volume should be around 8 cubic meters, with the rover-platform and the greenhouse as the dominating factors. The accumulated weight of all payloads will very likely not exceed 1 ton, leaving a generous 9 tons for the rover-platform.

A possible concern may lie in the technology readiness. We have brought up several payloads with novel and untested ideas, such as the mycotecture, balloons on Mars, snake-robot-drones or power generation utilizing gravity. Here, however, there are either groups already researching and developing these payloads, such as in the case of the mycotecture and snake-robots, or the technologies are very simple, as is the case for the balloon and gravity-power-winched-rover. The development of our rover should therefore not require until 2033 to conclude. Nevertheless, if it had to, it would involve around 600 people working on the project for 10 years. In terms of Full-Time Equivalent (FTE), this corresponds to 1,320,000 FTE required for that project, provided that one person per year represents on average roughly for 220 FTE.

We argue that our approach addresses all of the posed criteria through unique, creative, and valuable solutions. These range from those that explore Mars down from the deepest and most ancient caves, all the way to those soaring high up across its crimson skies. All while keeping a lookout for life that may have been or still inhabits the planet, but also for the life that we will bring and the life that might one day be on Mars. With that we are trying to answer the all-important question whether we can grow and harvest not only food in the form of plants, but also minerals produced by fungi and possibly even create entire habitats using the mycotecture.

We are doing all of that whilst trying to be resourceful and seeking to exploit every potential advantage, including the use of gravity for power generation, and equipping observational balloons with the means of turning wind into electrical energy. To finally tie it all up, these systems all fit-in together to work synergistically with each other to yield the highest benefit for us and our interplanetary future!